\documentclass[aps, prl, superscriptaddress, nofootinbib, preprintnumbers,twocolumn]{revtex4}
\usepackage{amsmath}
\usepackage{graphicx}

\newcommand{\chushi}[1]{}

\begin{document}
 \preprint{MISC-2012-02}
 \title{{\bf Techni-dilaton at 125 GeV}
 \vspace{5mm}}
\author{Shinya Matsuzaki}\thanks{
      {\tt synya@cc.kyoto-su.ac.jp} }
      \affiliation{ Maskawa Institute for Science and Culture, Kyoto Sangyo University, Motoyama, Kamigamo, Kita-Ku, Kyoto 603-8555, Japan.}
\author{{Koichi Yamawaki}} \thanks{
      {\tt yamawaki@kmi.nagoya-u.ac.jp}}
      \affiliation{ Kobayashi-Maskawa Institute for the Origin of Particles and 
the Universe (KMI) \\ 
 Nagoya University, Nagoya 464-8602, Japan.}
\date{\today}

\begin{abstract}
Walking technicolor predicts a light composite scalar, techni-dilaton, 
arising as a pseudo Nambu-Goldstone boson for the approximate scale symmetry spontaneously broken by techni-fermion condensation. 
We show that a light techni-dilaton with mass of around 125 GeV can explain presently observed excesses particularly
in the di-photon decay channel at LHC.

\end{abstract}
\maketitle

\section{Introduction}

The origin of mass is one of the most intriguing quest in 
particle physics. It is explained in the standard model (SM) by presence of Higgs boson. 
The LHC has now started searching for the SM Higgs and recently reported the first hint 
toward discovery of a Higgs-like object
around the mass of 125 GeV~\cite{ATLAS1213,CMS1213}. 
Thus we are now coming into an  exciting period of particle physics.

The ATLAS~\cite{ATLAS1213} and CMS~\cite{CMS1213} experiments reported 
the observation of some excesses at around 125 GeV in di-photon and weak gauge boson 
decay channels by $\sim$ 5 fb$^{-1}$ data, corresponding to the statistical significance about 2.5 $\sigma$ in total. 
In the weak gauge boson channels ($WW^* \to 2l 2 \nu$ and $ZZ^* \to 4 l$) 
the excessive signals around 125 GeV are compatible with the expected backgrounds~\cite{WW,ZZ}.    
In the di-photon channel, on the other hand, 
the excess around 125 GeV is about 3 $\sigma$ in the local significance level, 
and the signatures denoted by $\sigma \times {\rm BR}$ (cross section times branching ratio) 
are about 4 times larger than those of the SM Higgs resonance  at around 125 GeV, 
i.e, $(\sigma \times {\rm BR})_{\rm obs} \simeq 4 (\sigma \times {\rm BR})_{\rm SM-Higgs}$~\cite{diphoton},  
which would imply new physics beyond the SM.

Technicolor (TC)~\cite{Weinberg:1975gm,Farhi:1980xs} is 
an attractive idea to explain the origin of mass without introduction of fundamental Higgs boson, 
in a way that the electroweak symmetry is broken by the techni-fermion condensation 
just like the quark condensation in QCD. 
Although the original TC was ruled out long time ago by the excessive flavor-changing neutral currents (FCNC),
the walking TC (WTC) is a viable model beyond the SM, 
solving the FCNC problem by a large anomalous dimension $\gamma_m \simeq 1$
in the approximately scale/conformal-invariant dynamics~\cite{Yamawaki:1985zg}. 
(See also similar works~\cite{Akiba:1985rr} subsequently done without concept of anomalous dimension and the scale/conformal invariance.)
In sharp contrast to the original TC of a simple QCD-scale-up, the WTC 
predicts a relatively light composite scalar, techni-dilaton (TD)~\cite{Yamawaki:1985zg,Bando:1986bg},  
a pseudo Nambu-Goldstone boson associated with the spontaneously broken 
approximate scale symmetry.  
Since the TD mass is expected to be somewhat lighter than those of other techni-hadrons  of order ${\cal O}({\rm TeV})$, 
the TD is anticipated to be discovered at LHC instead of the SM Higgs.~\footnote{As to the notorious TC problem of  S and T parameters, possible solutions were already suggested~\cite{Matsuzaki:2011ie, Campbell:2011iw}.
Particularly, the issue on the $S$ parameter may be resolved in the case of walking~\cite{Appelquist:1991is,Harada:2005ru}.  
Even if  WTC in isolation cannot overcome this problem, there still exist a possibility that the problem may be 
resolved in the combined dynamical system including the SM fermion mass generation such as the extended TC 
(ETC) dynamics~\cite{Dimopoulos:1979es}, 
in much the same way as the solution (``ideal fermion delocalization'')~\cite{Cacciapaglia:2004rb} in Higgsless models. 
}

Recently the TD signatures of TD at LHC were actually studied 
in Ref.~\cite{Matsuzaki:2011ie}, focusing on the heavy mass range above 200 GeV,  
the region extensively searched at LHC before the recent report in the end of the last year~\cite{ATLAS1213,CMS1213}. 
In Ref.~\cite{Matsuzaki:2011ie} it was concluded that in the 
typical one-family WTC model~\cite{Farhi:1980xs} 
the heavy TD with mass around $\sim$  600 GeV\footnote{
$M_{\rm TD} \simeq 500 - 600 {\rm GeV}$ for the typical one-family model was suggested~\cite{Yamawaki:2007zz}, 
based on various explicit calculations,  
which are not conclusive, however,  due to the respective uncertainties in those computations. 
More reliable calculations such as the lattice simulations will yield a conclusive answer.
}
will be seen through the decays to $WW/ZZ$ or $\gamma\gamma$, along with the gigantic enhancement clearly distinguishable 
from the SM Higgs signatures~\footnote{
Phenomenological arguments on the TD in comparison with  the recent LHC data were also done in  
slightly different contexts~\cite{Coleppa:2011zx,Campbell:2011iw}. 
See also~\cite{Goldberger:2007zk}. }.

In this article, we simply extend the previous analysis~\cite{Matsuzaki:2011ie} down to the lower mass region and
explore a light TD with mass of around 125 GeV and compare its signatures with  
the present LHC data on this low mass region. 
Surprisingly, we find that the light TD in  the one-family WTC models actually should have  
predicted the excesses around 125 GeV particularly  in the di-photon  
decay channel before the observation reports~\cite{ATLAS1213,CMS1213} came out.

The weak boson decay channels turn out to be 
as much as the expected backgrounds consistently with 
the present LHC results: 
The gluon-fusion production cross section gets enhanced by about factor 10,    
$\sigma_{\rm TD}/\sigma_{h_{\rm SM}} \sim (g_{\rm TD}/g_{h_{\rm SM}})^2|1 + 2 N_{\rm TC}|^2 \sim {\cal O}(10)$,  
involving the enhancement from extra techni-quark loop contributions which are somewhat compensated by  
the overall suppression by TD coupling at 125 GeV (See Table~\ref{tab:production}), where we study
$N_{\rm TC}=3, 4, 5, 6$ for $SU(N_{\rm TC})$ WTC.   
Moreover, the techni-quark loop corrections make relatively larger  
the branching ratios for ${\rm TD} \to gg$ so that the branching fraction for other modes become 
about 10\% of the SM Higgs case at 125 GeV (See Table~\ref{tab:BR}).  
Accordingly, the cross section times branching ratio turns out to be   
of order of the SM Higgs one.

This mechanism is operative for the di-photon channel as well, 
though it gets enhanced more from  electromagnetically-charged techni-fermion loops (See Table~\ref{tab:signal}).   
Thus the di-photon signal becomes larger than the SM Higgs case, 
to be comparable with the current ATLAS and CMS data~\footnote{
Similar enhancement on the di-photon channel was discussed in Ref.\cite{Cheung:2011nv} in terms of radion. }. 
This enhancement will be even more eminent 
only in the di-photon channel, 
when $N_{\rm TC}$ is increased, a clear distinction from the SM Higgs.
For explicit formula  see Ref. {\cite{Matsuzaki:2011ie}.

\section{Techni-dilaton coupling}

As was discussed previously in Ref.~\cite{Matsuzaki:2011ie}, the TD couplings to the SM particles 
are almost identical to those of the SM Higgs, except for two ingredients: 
The scale set by the TD decay constant $F_{\rm TD}$ 
instead of the electroweak scale $v_{\rm EW}$ for the SM Higgs and 
the gluon, and photon couplings depending highly on particle contents of models of WTC. 
The essential discrepancy between the TD and SM couplings is therefore set by 
the ratio, 
\begin{equation} 
  \frac{g_{\rm TD}}{g_{h_{\rm SM}}} = \frac{(3-\gamma_m) v_{\rm EW}}{F_{\rm TD}} 
  \,, \label{g}
\end{equation}
where  the electroweak scale is $v_{\rm EW} \simeq 246$ GeV and  
$\gamma_m$ stands for the anomalous dimension of techni-fermion bilinear and 
$\gamma_m\simeq 1$ for WTC.

The TD decay constant $F_{\rm TD}$ and TD mass $M_{\rm TD}$ are related to the vacuum energy 
density ${\cal E}_{\rm vac}=\langle \theta_\mu^\mu\rangle/4$ 
through partially conserved dilatation current for the trace anomaly: 
\begin{equation} 
F_{\rm TD}^2 M_{\rm TD}^2 
= - 4\,\langle \theta_\mu^\mu\rangle =-16\, {\cal E}_{\rm vac} 
\,, \label{PCDC}
\end{equation}
where $\theta_{\mu\nu}$ is the energy-momentum tensor. 
The vacuum energy density ${\cal E}_{\rm vac}$ is dominated by the  techni-gluon condensation 
induced by the loop of the  techni-fermion 
with 
dynamical mass $m_F$, which 
can be written in 
a generic manner as 
\begin{equation} 
 \langle  \theta_\mu^\mu \rangle = 4 {\cal E}_{\rm vac}  
  = - \kappa_V \left( \frac{ N_{\rm TC} N_{\rm TF} }{2 \pi^2}  \right) m_F^4 
  \,,  
  \label{Vac}
\end{equation} 
with $\kappa_V$ being the overall coefficient  
which is in principle calculable by the  nonperturbative analysis. 
$N_{\rm TF}$ denotes the flavor number of techni-fermions.

The dynamical techni-fermion mass $m_F$ can, on the other hand, be related to 
the techni-pion decay constant $F_\pi$:  
\begin{equation} 
  F_\pi^2 = \kappa_F^2 \frac{N_{\rm TC}}{4 \pi^2} m_F^2 
\,,   
\end{equation}
with the overall coefficient $\kappa_F$ and the property of $N_{\rm TC}$ scaling taken into account. 
The scale of $F_\pi$ is set by the electroweak scale $v_{\rm EW}$ along with $N_D$ as 
$F_{\pi} = v_{\rm EW}/\sqrt{N_D}$, where $N_D$ denotes the number of electroweak doublet techni-fermions.    
With these combined, one can express 
$F_{\rm TD} M_{\rm TD}$ in Eq.(\ref{PCDC}) 
in terms of $N_{\rm TC}, N_{\rm TF}$ 
and $\kappa_{V,F}$, once $F_{\pi} = v_{\rm EW}/\sqrt{N_D}$ is fixed.

 As was done in Ref.~\cite{Matsuzaki:2011ie}, 
the values of $\kappa_V$ and $\kappa_F$ may be quoted from the latest result~\cite{Hashimoto:2010nw} on a 
ladder Schwinger-Dyson analysis for a modern version of WTC~\cite{Lane:1991qh,Appelquist:1996dq, Miransky:1996pd}: 
\begin{equation} 
 \kappa_V \simeq 0.7 \,, \qquad 
\kappa_F \simeq 1.4 
\,.  \label{kappas}
\end{equation} 
In that case $N_{\rm TF}$ is fixed by the criticality condition for the walking regime as~\cite{Appelquist:1996dq} 
\begin{equation} 
 N_{\rm TF} \simeq 4 N_{\rm TC}
 \,,
 \label{criticality}
\end{equation} 
where $N_{\rm TF} = 2 N_D + N_{\rm EW-singlet}$, 
with $N_{\rm EW-singlet}$ being the number of the  electroweak/color-singlet techni-fermions, ``dummy'' techni-fermions
introduced in order to fulfill the criticality condition, which serve to reduce the TD coupling $g_{\rm TD}$ by enhancing  
$F_{\rm TD}$ through Eqs.(\ref{PCDC}) and (\ref{Vac}). 
 Taking the original one-family model~\cite{Farhi:1980xs} with $N_{D} = 4$~as a definite benchmark, 
we thus evaluate $m_F$, $F_{\rm TD}$ and $g_{\rm TD}/g_{h_{\rm SM}}$ in Eq.(\ref{g}) to get 
\begin{eqnarray} 
 && 
m_F \simeq 319 \, {\rm GeV} \sqrt{ \frac{3}{N_{\rm TC}}} \,, \quad 
 F_{\rm TD} \simeq 1836 \, {\rm GeV} \left( \frac{125\,{\rm GeV}}{M_{\rm TD}}  \right)  
 \nonumber \\ 
&& 
   \frac{g_{\rm TD}}{g_{h_{\rm SM}}} \simeq 0.27  \left( \frac{M_{\rm TD}}{125\,{\rm GeV}} \right) 
\,. \label{vals}
\end{eqnarray}
Note that $F_{\rm TD}$ and hence the TD coupling is independent of $N_{\rm TC}$ when $N_{\rm TF} \simeq 4 N_{\rm TC}$ is used. 
The plot of $g_{\rm TD}/g_{h_{\rm SM}}$ as a function of $M_{\rm TD}$ is 
shown in Fig.~\ref{g-ratio-LSD}.~\footnote{
At this point we may remark on  stability of the light TD mass against radiative corrections.  
As a pseudo Nambu-Goldstone boson of scale invariance the quadratic divergence is suppressed by the 
scale invariance for the walking energy region $m_F < \mu <\Lambda$, where $\Lambda$ is the intrinsic scale of the walking TC, roughly taken as the order of the ETC scale $\Lambda_{\rm ETC}$.
The scale symmetry breaking in the ultraviolet region $\mu>\Lambda$ has no problem for the naturalness as usual like in the QCD and the QCD-scale-up TC where the  theory has  only logarithmic divergences. 
Only possible source of the scale symmetry violation is from $\mu<m_F$,  giving  rise to the quadratically divergent corrections $\delta M_{\rm TD}^2 \sim  \mu^2/(4\pi)^2< m_F^2/(4\pi)^2$, which is evaluated from Eq. (\ref{vals}) 
as only 2 percent corrections to $M_{\rm TD} (\simeq 125$ GeV). Higher loop corrections are even more dramatically suppressed by powers of  $(m_F/(4\pi F_{\rm TD}))^2$.
}

 \begin{figure}
\begin{center} 
\includegraphics[scale=0.55]{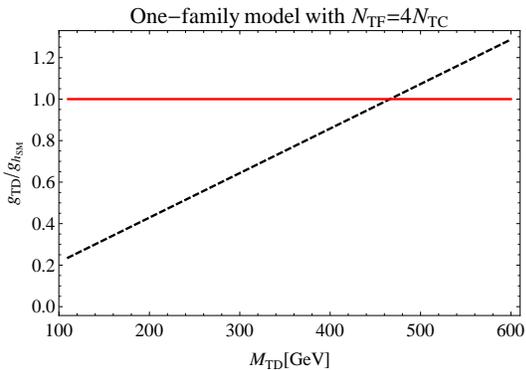}
\end{center}
\caption{ $g_{\rm TD}/g_{h_{\rm SM}}=(3-\gamma_m)v_{\rm EW}/F_{\rm TD}$ with $\gamma_m\simeq 1$, 
with respect to the TD mass $M_{\rm TD}$ in a range from 110 to 600 GeV for one-family models with 
$N_{\rm TF}=4 N_{\rm TC}$ fixed. 
}  
\label{g-ratio-LSD}
\end{figure}

\section{The LHC signatures at  125 GeV}

 Using the values in Eq.(\ref{vals}) and formulas~\footnote{
 For the $WW^*$ and $ZZ^*$ decays we have quoted the corresponding formulas for the SM Higgs given in Ref.~\cite{Spira:1997dg} just simply 
 by replacing $v_{\rm EW}$ with $F_{\rm TD}/2$. 
} previously reported in Ref.~\cite{Matsuzaki:2011ie}, 
we compute the TD LHC production cross section times branching ratios normalized to 
the corresponding quantities for the SM Higgs. 
Here we focus on the one-family model with $N_{\rm TC}=3, 4, 5, 6$.  
The SM Higgs branching ratios and LHC production cross sections at 7 TeV are read off 
from Ref.~\cite{Dittmaier:2011ti}.

The production cross section is highly dominated by the gluon fusion 
process since the TD couplings to the weak bosons and fermions are suppressed just by amount of 
$(g_{\rm TD}/g_{h_{\rm SM}})^2 = {\cal O}(10^{-2})$  
for the mass region we are interested in, around 125 GeV (See Eq.(\ref{vals}) and Table~\ref{tab:production}). 
The gluon fusion production, on the other hand, gets enhanced due to the presence of techni-quarks 
carrying the QCD color.

The same argument is applicable to the branching fraction as well: 
The di-gluon decay channel becomes fairly enhanced in the branching fraction to highly exceed 
the $b \bar{b}$ channel (See Table~\ref{tab:TDBR}),  
so that the other decay channels are relatively suppressed compared to the SM Higgs case (See Table~\ref{tab:BR}).       
The total width of TD at around 125 GeV is however as small as the SM Higgs one $\Gamma_{\rm tot}^{\rm TD}(125\,{\rm GeV})\sim$ a few MeV.

\begin{table}[h] 
\begin{tabular}{|c|c|c|c|} 
\hline 
\hspace{5pt} $N_{\rm TC}$ \hspace{5pt} 
&\hspace{5pt} $g_{\rm TD}/g_{h_{\rm SM}}$ \hspace{5pt} 
&\hspace{5pt} $\sigma_{\rm TD}/\sigma_{h_{\rm SM}}|_{\rm GF}$ \hspace{5pt}
&\hspace{5pt} $\sigma_{\rm TD}/\sigma_{h_{\rm SM}}|_{\rm VBF}$  \hspace{5pt}  \\ 
\hline \hline 
3 & 0.27 & 3.8 & 0.072 \\ 
4 & 0.27 & 6.3 & 0.072  \\ 
5 & 0.27 & 9.4 & 0.072  \\ 
6 & 0.27 & 13  & 0.072 \\ 
\hline 
\end{tabular}
\caption{
The estimated numbers at $M_{\rm TD}=125$ GeV relevant to the TD LHC production processes at 7 TeV, 
compared with the corresponding ones for the SM Higgs. GF and VBF label gluon and vector boson fusions, respectively.    
}\label{tab:production} 
\end{table}

\begin{table}[h] 
\begin{tabular}{|c|c|c|c|} 
\hline 
\hspace{15pt} $N_{\rm TC}$ \hspace{15pt} 
&\hspace{15pt} BR$_{gg}$ \hspace{15pt} 
&\hspace{15pt} BR$_{b \bar{b}}$ \hspace{15pt}
&\hspace{15pt} others  \hspace{15pt} \\ 
\hline \hline 
3 & 82\% & 10\% & 8\% \\ 
4 & 88\% & 7\% & 5\%  \\ 
5 & 92\% & 5\% & 3\%  \\ 
6 & 94\% & 4\%  & 2\% \\ 
\hline 
\end{tabular}
\caption{
The TD branching fraction at $M_{\rm TD}=125$ GeV. 
}\label{tab:TDBR} 
\end{table}

\begin{table}[h] 
\begin{tabular}{|c|c|c|c|} 
\hline 
\hspace{15pt} $N_{\rm TC}$ \hspace{15pt} 
&\hspace{15pt}  $r_{\rm BR}^{2\gamma}$ \hspace{15pt} 
&\hspace{15pt}  $r_{\rm BR}^{2g}$ \hspace{15pt} 
&\hspace{15pt} $r_{\rm BR}^{\rm others}$  \hspace{15pt} \\ 
\hline \hline 
3 & 0.079 & 10 & 0.19 \\ 
4 & 0.18 & 10 & 0.12  \\ 
5 & 0.26 & 11 & 0.086  \\ 
6 & 0.33 & 11 & 0.063 \\ 
\hline 
\end{tabular}
\caption{
The TD branching fraction at $M_{\rm TD}=125$ GeV compared with the SM Higgs, 
$r_{\rm BR}^{X}\equiv {\rm BR}({\rm TD} \to X)/{\rm BR}(h_{\rm SM} \to X)$. 
The label ``others" denotes other decaying particles relevant to 
this mass range, such as 
$WW^*, ZZ^*, b\bar{b}, c \bar{c}, \tau^+ \tau^-$.  
}\label{tab:BR} 
\end{table}

\begin{table}[h] 
\begin{tabular}{|c|c|c|c|} 
\hline 
\hspace{15pt} $N_{\rm TC}$ \hspace{15pt} 
&\hspace{15pt}  $R_{2\gamma}$ \hspace{15pt} 
&\hspace{15pt}  $R_{2g}$ \hspace{15pt} 
&\hspace{15pt} $R_{\rm others}$  \hspace{15pt} \\ 
\hline \hline 
3 &0.28 & 35 & 0.67  \\ 
4 & 1.0 & 63 & 0.72   \\ 
5 & 2.3 & 99 & 0.75  \\ 
6 & 4.1 & 142 & 0.77 \\ 
\hline 
\end{tabular}
\caption{
The TD signatures at $M_{\rm TD}=125$ GeV normalized to those of the SM Higgs, 
$R_{X} \equiv  \sigma_{\rm TD} \times {\rm BR}({\rm TD} \to X)/[ \sigma_{h_{\rm SM}} \times {\rm BR}(h_{\rm SM} \to X)]$, 
where $\sigma_{i}= \sigma_{i}|_{\rm GF} + \sigma_{i}|_{\rm VBF}$ ($i={\rm TD}, h_{\rm SM}$). 
The label ``others" means the same as in Table~\ref{tab:BR}.  
}\label{tab:signal} 
\end{table}

The result on the TD signatures at 125 GeV is summarized in 
Table~\ref{tab:signal}.  
 We see that the di-photon signal is fairly sensitive to the number of $N_{\rm TC}$:  
When $N_{\rm TC}=6$ it is 
close to the amount of the presently observed excess $\sim 4 \times \sigma_{\rm h_{\rm SM}} 
\times {\rm BR}(h_{\rm SM} \to \gamma\gamma)$~\cite{diphoton}, while 
it exceeds the present observation for $N_{\rm TC}\ge 7$. 
 One can understand this feature by considering a ratio $R_{2\gamma}/R_{WW/ZZ}$ whose 
$N_{\rm TC}$-dependence can be roughly described numerically 
$(R_{ 2\gamma}/R_{WW/ZZ})|_{N_{\rm TC}} \sim (1 + 0.3 N_{\rm TC})^2$ at $M_{\rm TD}=125$ GeV. 
The di-photon excess therefore grows even more as $N_{\rm TC}$ is
increased.  It is sharply contrasted to other channels 
 including the weak boson decay channels which are almost insensitive 
to $N_{\rm TC}$, staying in the range consistent with the present data on the weak boson decay channels~\cite{WW,ZZ} 
as well as the fermionic modes~\cite{tau}.

 To be more explicit, in Fig.~\ref{TD-WW-ZZ-diphoton} we plot the TD signatures as a function of 
$M_{\rm TD}$ varied from 110 to 150 GeV, 
along with the current ATLAS and CMS 95\% C.L. upper limits on 
$WW^*, ZZ^*, \gamma\gamma$ channels and their expected backgrounds~\cite{WW,ZZ,diphoton}.

The estimated signals for the weak boson channels can be pulled up by about 30\% 
($R_{WW/ZZ} \simeq 0.77 \to 1.0$ at 125 GeV when $N_{\rm TC}=6$) 
to be within a range consistent with the expected backgrounds for the weak boson channels~\cite{WW,ZZ}: 
This error comes from a theoretical uncertainty associated with 
the estimate 
 of $\kappa_V$ and $\kappa_F$ in Eq.(\ref{kappas}), 
arising from the deviation of the criticality condition~\cite{Hashimoto:2010nw}: 
$\kappa_F  \simeq 1.4 \to 1.49$ (shift by about 6\%), $\kappa_V \simeq 0.7 \to 0.81$ (shift by about 14\%) 
at the criticality. The expected uncertainty about $(g_{\rm TD}/g_{h_{\rm SM}})^2$ will be about 30\%.  
Similar improvement can be made for the fermionic modes~\cite{tau} as well, so that 
all the signatures other than the di-photon channel will be consistent with the expected backgrounds at about $2\sigma$ level.  
Thus the excess of only the di-photon channel will be a salient feature of the TD discriminated from the SM Higgs.

 \begin{figure}[h]
\begin{center}
      \includegraphics[scale=0.55]{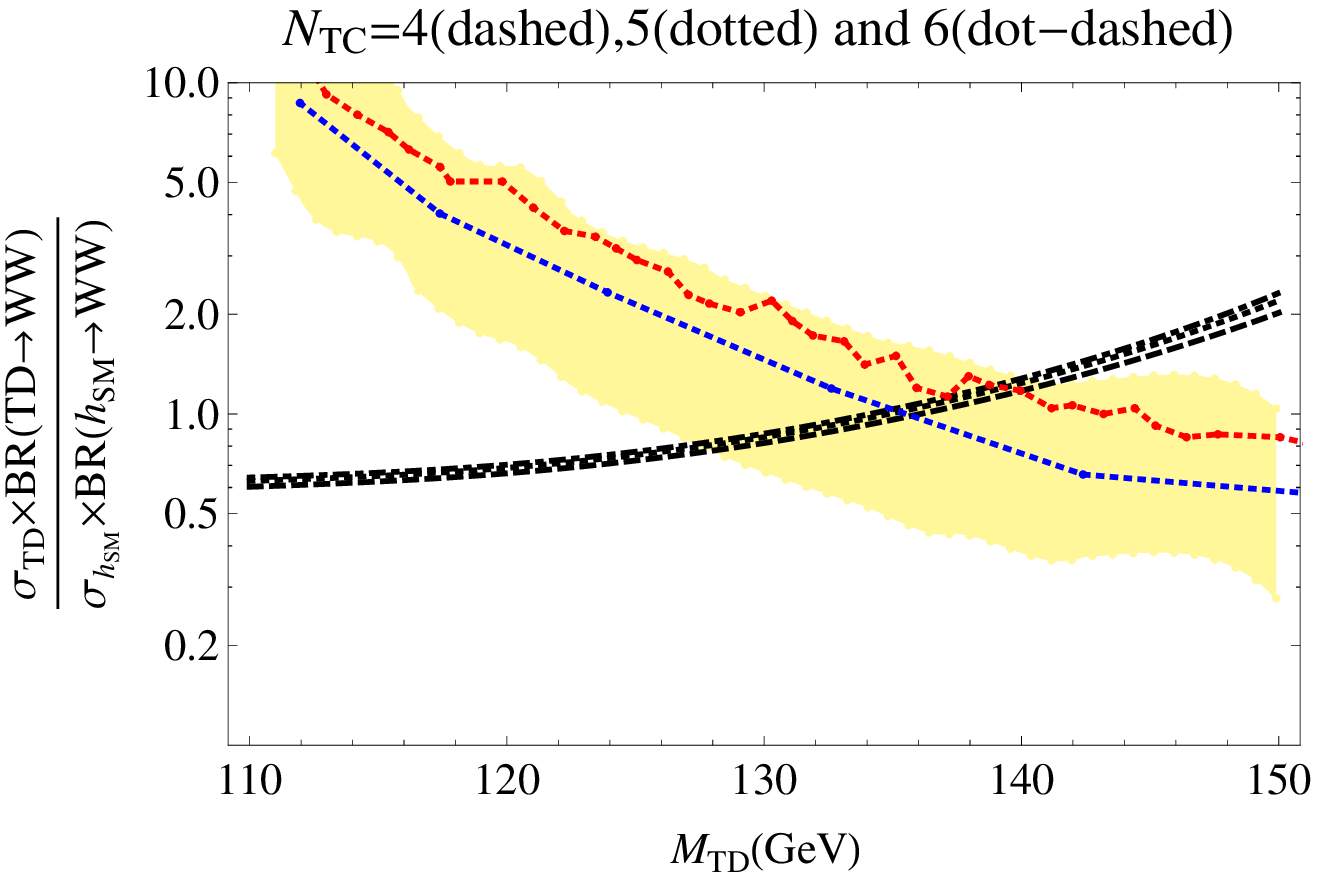}
      \includegraphics[scale=0.55]{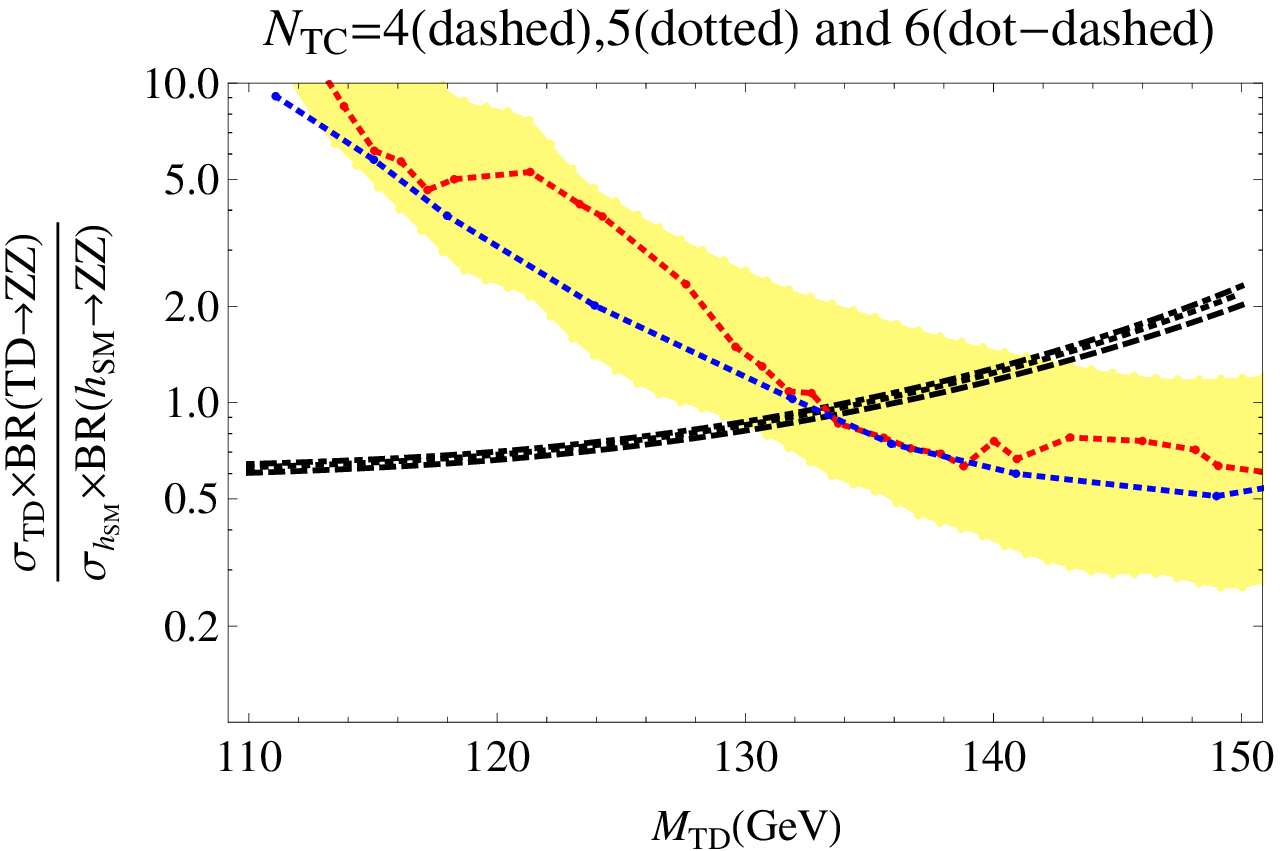}
      \includegraphics[scale=0.52]{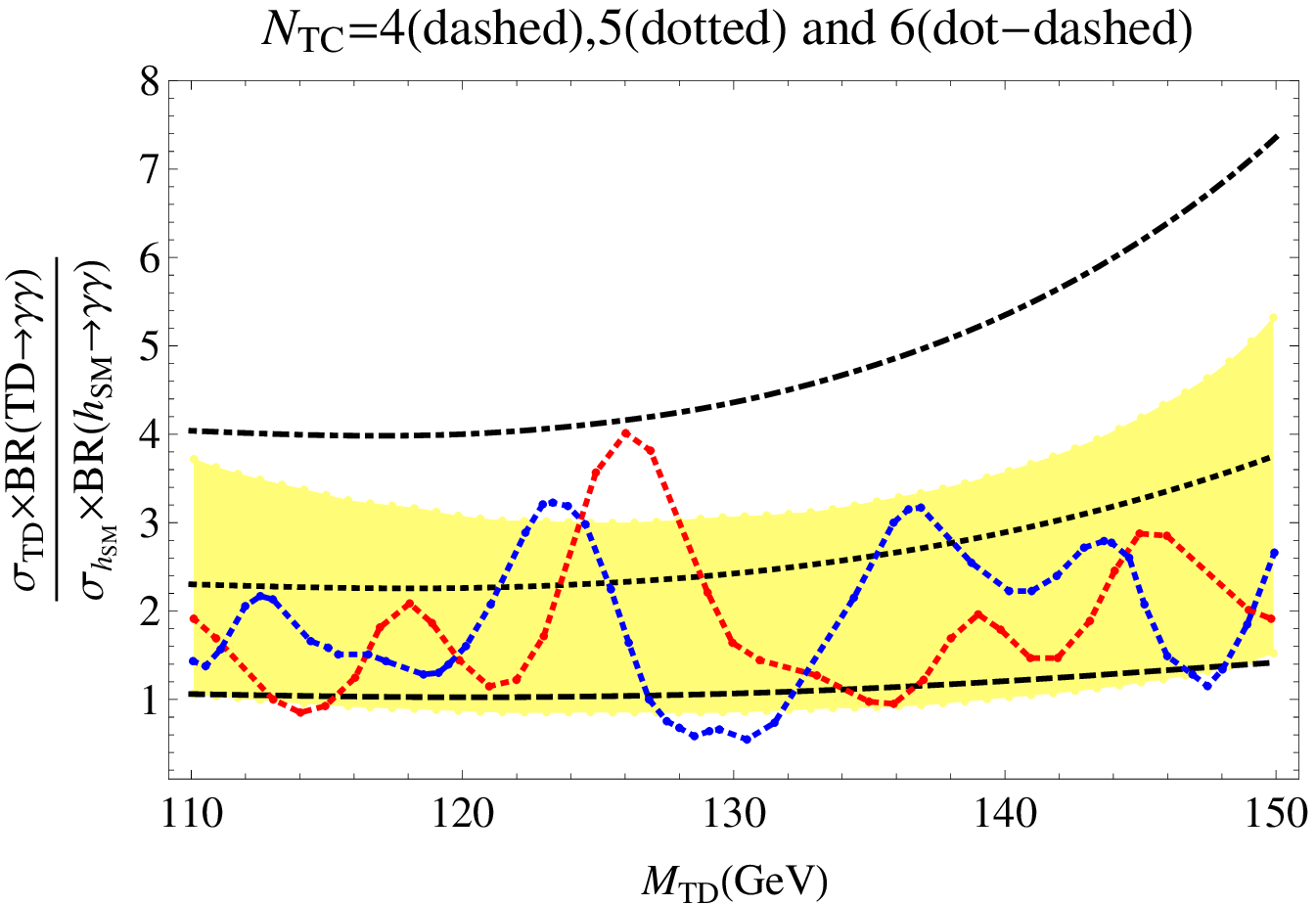}
\caption{ 
The TD LHC cross section at 7 TeV times branching ratios for $WW^*$ (top panel), $ZZ^*$ (middle panel) and 
di-photon (bottom panel) decay channels  
normalized to the corresponding quantities for the SM Higgs, in comparison with the current 
ATLAS (red curves) and CMS (blue curves) data~\cite{WW,ZZ,diphoton} at 95\% C.L.for the low Higgs mass range. 
The yellow band denotes the expected backgrounds within 2 $\sigma$ level.   
\label{TD-WW-ZZ-diphoton}
}
\end{center} 
 \end{figure}

\section{Conclusion}

 To conclude, we have explored a light TD with mass of around 125 GeV and compare its signatures with  
the present LHC data available for this low mass region. 
We showed that the light TD in the one-family WTC models actually gives the signals consistent with 
the presently observed excesses around 125 GeV particularly in the di-photon channel. 
 The main results 
 in Fig.~\ref{TD-WW-ZZ-diphoton} shows that 
 when $N_{\rm TC}$ increases, only the di-photon channel excess grows, while other channel stay unchanged.
 This is a clear distinction from the SM Higgs.
 Then,  if the excessive di-photon signals develop at the upcoming experiments to reach 
the desired significance level, while other channels like the weak boson signals essentially stay at  
the present significance, it  would imply  
the discovery of the 125 GeV TD.

\section*{Acknowledgments}

This work was supported by 
the JSPS Grant-in-Aid for Scientific Research (S) \#22224003 and (C) \#23540300 (K.Y.).

\end{document}